\documentclass{PoS}

\usepackage{graphicx}

\def\svev#1{\left\langle #1\right\rangle}       

\def\Tr{{\rm Tr}\,}

\def\sl#1{\rlap{\hbox{$\mskip 1 mu /$}}#1}      

\newcommand{\bee}{\begin{equation}}
\newcommand{\ee}{\end{equation}}
\newcommand{\beea}{\begin{eqnarray}}
\newcommand{\eea}{\end{eqnarray}}

\title{Composite phenomenology as a target for lattice QCD}

\ShortTitle{Composite phenomenology as a target for lattice QCD}

\author{\speaker{Thomas DeGrand}\\
Department of Physics,
University of Colorado, Boulder, CO 80309, USA\\
E-mail: \email{thomas.degrand@colorado.edu}}

\author{Ethan T.~Neil \\
Department of Physics, University of Colorado, Boulder, CO 80309, USA \\
RIKEN-BNL Research Center, Brookhaven National Laboratory, \\ Upton, New York 11973, USA\\
E-mail: \email{ethan.neil@colorado.edu}}

\abstract{
Some recent beyond Standard Model phenomenology is based on new strongly interacting dynamics
of $SU(N)$ gauge fields coupled to various numbers of fermions. When $N=3$ these systems are
analogues of QCD, although the fermion masses are typically different from  -- and heavier than --
 the ones of real world QCD.
Many quantities  needed for phenomenology from these
models have been computed on the lattice. We are writing a guide for these
phenomenologists, telling them about lattice results. We'll tell you (some of) what they are
interested in knowing.
}

\FullConference{The 36th Annual International Symposium on Lattice Field Theory - LATTICE2018\\
                22-28 July, 2018\\
                Michigan State University, East Lansing, Michigan, USA.}

\begin{document}


Phenomenologists continue to construct and analyze a wide variety of theories for physics beyond the 
Standard Model (BSM). Some of these theories contain nonperturbative systems --
non-Abelian gauge fields coupled to fermions (and perhaps scalars) in which the
gauge symmetry is unbroken and the gauge dynamics are confining. Many of these systems
are accessible to lattice simulation. A subset of them are so accessible to lattice simulation that
they have (probably) already been simulated -- the gauge group is $SU(3)$ and the matter
content is a set of
fundamental representation fermions. What makes them different from real world QCD is that generally
 the pseudoscalar to vector meson mass ratio is bigger than in Nature.  The corresponding lattice data 
 for QCD at unphysically heavy quark masses exists, but it is generally thought of merely an intermediate result
  on the way to the QCD physical point.  This means that although such lattice results could have an impact in BSM phenomenology, they are (by and large) not presented in a way which is accessible to researchers outside the lattice community for such purposes.
  
We are trying to collect lattice data which might impact beyond-Standard-Model phenomenology
and present it to the community working in that area \cite{TDENprep}.
You lattice people are not really the audience for the paper we want to write.
However, it's your data we are trying to collect. We may not have found the best examples
of the things we want to show. 
The purpose of this writeup is to tell you, a lattice QCD practitioner,
 what selected phenomenologists have told us they are interested in,
and to show you some of what we have found.


Our focus is on BSM models including a non-perturbative sector that resembles ``heavy QCD'': SU$(3)$ with fundamental-irrep fermions heavier relative to the confinement scale than the light quarks of real-world QCD.  Generically, these systems are examples of ``hidden valleys'' -- new confining sectors
with some weak coupling to the visible sector of the Standard Model. Early representative
examples include Refs.~\cite{Strassler:2006im,Pospelov:2007mp,Han:2007ae,Craig:2015pha}.
Some constructions are pointed at the hierarchy problem; some at dark matter.
 Quantities which appear most often in them, which might
be lattice targets are: 
\begin{itemize}
\item spectroscopy (of course) -- but some translation of scales from QCD is required,
\item  the pseudoscalar decay constant and
other parameters of the low energy chiral effective theories (most useful for extracting
Higgs properties from the low energy effective field theory),
\item decay constants, representing the matrix element of some strong bound state to ``vacuum''; these are
 necessary to describe the decay of bound states through other interactions, e.~g.~a Z-boson or a dark photon.
  Certain decay constants, such as the vector (and perhaps axial vector), also appear in phenomenological 
descriptions such as vector meson dominance which are prevalent in the strongly-coupled BSM literature.
\item other simple matrix elements, for example matrix elements of the scalar current, which determine the 
Higgs boson coupling to the new physics sector.  An example of this coupling is the nuclear sigma term (which
 describes the coupling of the Higgs to the nucleon in a direct detection dark matter decay amplitude). 
If dark matter is composite
there is the analog matrix element coupling the Higgs to some scalar dark current.
\end{itemize}

These are very general lists, encompassing much of what is done in lattice QCD.  Of course, that is the point; 
most existing lattice QCD calculations, particularly at heavy fermion mass, already contain results which can
 be of interest for BSM phenomenology - so long as they are presented in the right way!  We now move on to two
 more concrete examples and discuss where lattice results could be most impactful, beginning with ``twin Higgs''.


Little beyond-Standard-Model phenomenology directed at solving the hierarchy problem
involves QCD-like dynamics. One exception is the twin Higgs model which introduces
a copy of every Standard Model fermion, interacting with a new strong $SU(3)$ 
gauge symmetry with a different
confinement scale.
The original reference is \cite{Chacko:2005pe} and recent papers
 are \cite{Craig:2015pha,Cheng:2015buv,Chacko:2018vss}. $SU(3)$ is mandatory for graph cancellation in loops
 between Standard Model
particles and their twins. Otherwise, there are many variant models.
 
 The original reference \cite{Chacko:2005pe}  has a copy of every Standard Model fermion.
 The twin fermion masses are different than the Standard Model ones
 because the Yukawa couplings are different.  Obviously, properties of the 
 strongly interacting twin sector are those of full QCD, but with  different  fermion masses.
Later papers remind us that
 having many  more light particles
 than are already in the Standard Model is bad for nucleosynthesis, 
so the full twin scenario seems to be disfavored
 (unless the scale of the twin sector is very high, or unless all the light constituents
 can decay to Standard Model particles before nucleosynthesis).
 
 The next set of models restrict the twin quarks only to be partners of the top and bottom quarks.
 The justification for doing this  is that the top
 quark has the biggest Yukawa coupling and is the biggest player in the hierarchy problem, the quadratic
 dependence of the Higgs mass on higher new physics scales.
 It is reminiscent of partial compositeness, where the quarks get their mass by mixing with some composite operator.
 Most of the phenomenology of this scenario restricts itself to the top and bottom quark doublet.

 Several groups, including
Refs.~\cite{Craig:2015pha,Cheng:2015buv} have written about this scenario. 
 With $b$ quarks heavier than the scale of glueball bound states, the spectroscopy is 
very different from real world QCD:
 there are glueballs, which are basically quenched glueballs, and there are $\bar b b$
 quarkonia, basically quenched quarkonia. (The mirror top quarks decay as in the real world.)
The quarkonia  can only decay by glueball emission.
 
The lattice literature on quarkonia does not seem to have affected twin - related
 phenomenology. We could not find any lattice spectroscopy away from the physical
$c$ and $b$ masses. This would be interesting (and trivial to do, if you have the code).
Phenomenologists know about quenched lattice glueball spectroscopy and cite 
Refs.~\cite{Morningstar:1999rf,Chen:2005mg}.
 They care about the coupling of glueballs to quarkonia. 
The one lattice paper on this we know is Ref.~\cite{Lee:1999kv}. A quick glance
 does not reveal many citations to this by modern BSM phenomenologists.
 It's an old paper -- can one do better?

 
   Our second example comes from dark matter.  At present there are more lattice targets in dark matter phenomenology.
Ref.~\cite{Kribs:2016cew} is a survey of confining systems which have a place there.
  
Self interacting dark matter is characterized by having a dominant
decay process $3\rightarrow 2$. Some 
phenomenology assumes that the dark matter is bound states.
In \cite{Hochberg:2014dra,Hochberg:2014kqa} the dark matter is
the pions of a hidden sector.
A recent proposal is \cite{Berlin:2018tvf}.
It is an explicit model with
 $SU(3)$ gauge dynamics with $N_f=3$ light flavors, but (unlike QCD) it has  $m_{PS}/m_V \sim 1/2$.
The authors are interested in computing the $3\rightarrow 2$ amplitude in this system.
They are also interested in properties of their vector meson. When a ``dark photon'' is included in the dark sector, it mixes not only with the ordinary photon via a term in the Lagrangian (${\cal L}_I \sim \epsilon B_{\mu \nu} F_{\mu \nu}$), but also with the equivalent of the rho meson.  This mixing involves the vector meson decay constant
($f_V$ in QCD language), which, to set conventions, we define as
$
\langle 0| \bar u \gamma_i d  | V\rangle = m_V^2 f_V \epsilon_i .
$

Workers in this genre compute these quantities using a phenomenological effective Lagrangian,
basically the usual chiral Lagrangian for the Goldstones augmented by extra vector meson fields.
\bee
{\cal L} = \frac{F^2}{4}\Tr (D_\mu U D^\mu U^\dagger) - \frac{1}{8} \Tr G_{\mu\nu}G^{\mu\nu} + \dots
\ee
which is built of the Goldstone field
$U= \exp(i\Phi/F)$ and vector mesons $V_\mu$ introduced via covariant derivative
\bee
D_\mu \Phi = \partial_\mu \Phi + \frac{ig}{2}[\Phi,V_\mu] .
\ee
They have a self coupling from
$
G_{\mu\nu}=\partial_\mu V_\nu - \partial_\nu V_\mu .
$ 
The 
$\dots$ in ${\cal L}$ includes phenomenological $V$ mass terms, couplings, and so on.
The acronym ``KSRF'' 
(Kawarabayashi, Suzuki, Riadzuddin, Fayazuddin
\cite{Kawarabayashi:1966kd,Riazuddin:1966sw}) labels results from these  models.
Ref.~\cite{Klingl:1996by} is a  survey of them.

 Phenomenologists get their $3\rightarrow 2$
vertices out of a combination of the Wess-Zumino-Witten term in a chiral Lagrangian, and the
coupling of a vector meson to pseudoscalars, $g_{VPP}$.
They infer  $f_V$ and $g_{\rho \pi\pi}$ from the KSRF relations,
\bee
f_V = \sqrt{2} \frac{f_{PS}}{M_V}.
\label{eq:ksrfv}
\ee
and
\bee
g_{VPP} = \frac{M_V}{f_{PS}}.
\label{ksrfg}
\ee
In these conventions the vector meson decay width is
\bee
\Gamma(V \rightarrow PP) \simeq \frac{g^2_{VPP} }{48 \pi m_V^2}(m_V^2-4m_{PS}^2)^{3/2} 
\label{eq:vector_width}
\ee
and
\bee
\Gamma(V \rightarrow e^+ e^-) = \frac{4\pi\alpha^2}{3} m_V f_V^2 \svev{q}^2
\ee
where $\svev{q}$ is the average quark charge in the valence wave function.

\begin{figure}
\begin{center}
\includegraphics[width=0.9\textwidth,clip]{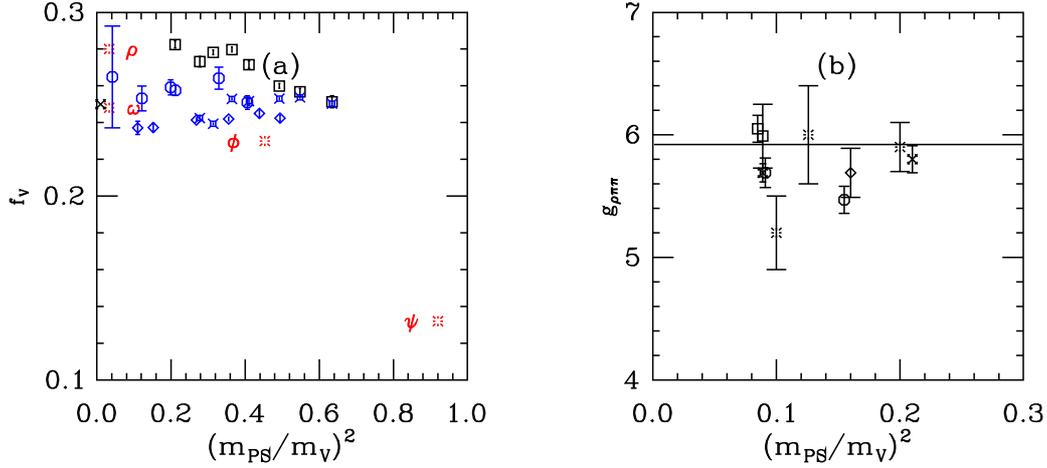}
\end{center}
\caption{ 
KSRF physics from the lattice:
(a) Vector meson decay constant 
$f_V$ versus
$(m_{PS}/m_V)^2$.
Squares: direct lattice calculations of $f_V$; blue symbols $f_V$ from KSRF $f_{PS}$ and $m_V$.
The fancy cross is the KSRF result for massless quarks
from the physical rho mass and pion decay constant. Results for physical particles are shown in red.
(b) The vector meson decay constant $g_{VPP}$ from lattice calculations,
 as a function of $(m_{PS}/m_V)^2$.
 Symbols are
squares, Ref.~\cite{Bulava:2016mks} and \cite{Bulava:2017stw};
fancy crosses, Ref.~\cite{Dudek:2012xn} and \cite{Wilson:2015dqa};
octagons, Ref.~\cite{Guo:2016zos};
diamond, Ref.~\cite{Alexandrou:2017mpi}
and bursts, Ref.~\cite{Erben:2017hvr}.
The line is the KSRF relation with physical values for the rho mass and $f_\pi$.
\label{fig:ksrf}}
\end{figure}

Does this phenomenology produce reasonable results? The situation for $f_V$ is shown in panel (a) of
Fig.~\ref{fig:ksrf}, showing direct lattice results, KSRF predictions from lattice data, and
experimental results (from radiative decays of vector mesons), and the KSRF 
relation from the real world
$m_\rho$ and $f_{PS}$. Yes, phenomenology works.

There are now many direct lattice calculations of $g_{VPP}$ from simulations
 in finite volume a la L\"uscher.
Lattice data from several groups is displayed in panel (b) of Fig.~\ref{fig:ksrf}, along with
the KSRF relation itself, evaluated using the physical values of $m_V$ and $f_{PS}$.
The agreement of lattice data with the relation is again excellent.

Finally, we briefly discuss scale setting.  Lattice calculations produce only dimensionless ratios; some physical quantity must be chosen to remove the lattice spacing dependence from these ratios and present final results in physical units like GeV.  The same 
procedure is needed to use lattice QCD results in the context of BSM phenomenology, but in this context the
 physical units will be different, and often variable over a wide range depending on the model parameters. 
 To allow scale setting for phenomenology, it is crucial to present intermediate results, i.~e.~ratios of 
physical quantities, in addition to final results in units of GeV.  Moreover, the most useful ratio for scale
 setting may be different depending on the BSM model; for example, in dark matter models the mass of the
 dark matter candidate bound state is a natural choice for scale setting.

\begin{figure}
\begin{center}
\includegraphics[width=0.5\textwidth,clip]{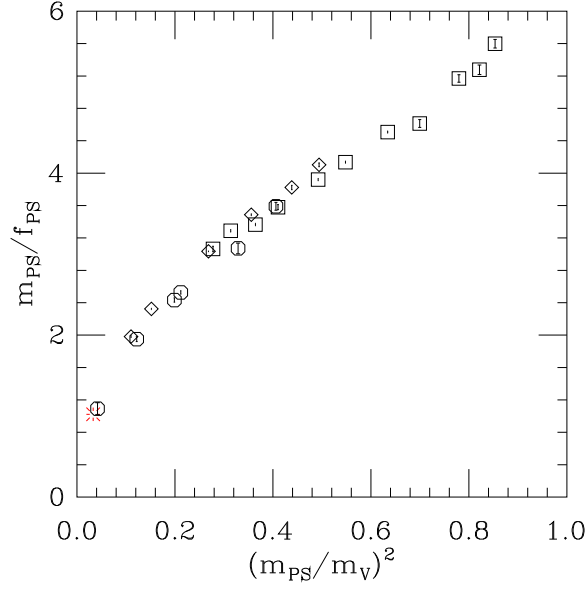}
\end{center}
\caption{Ratio of pseudoscalar mass to decay constant  as a function of $(m_{PS}/m_V)^2$.
Data are octagons from Ref.{\protect{\cite{Aoki:2008sm}}},
diamonds from Ref.~{\protect{\cite{WalkerLoud:2008bp}}},
and squares from a large statistics follow-on to Ref.~{\protect{\cite{DeGrand:2016pur}}}.
In our convention the physical $f_\pi=130$ MeV, shown as the red burst.
\label{fig:mpifpi}}
\end{figure}

One example of a broadly useful and lattice-accessible ratio is the quantity $m_{PS}/f_{PS}$.
It sometimes appears as a free parameter
in the phenomenological literature, where it is allowed to vary over a large
range.
(For example, see Fig.~1 of Ref.~\cite{Hochberg:2014kqa}. This is not $SU(3)$, but that is not important
for the point we are about to make.)  In QCD, simulations show that this ratio is always smaller than about 5-6.
A compilation of lattice data is shown in Fig.~\ref{fig:mpifpi}.
The range is even smaller if one wants to be in the chiral regime ($m_{PS}/m_V$ small).
Knowing that simulations can bound the possible ranges of quantities like $m_{PS}/f_{PS}$ can sharpen
 phenomenological predictions.

Of course, there are many other things to say, but we just
conclude with two remarks:

First, there is a market for $SU(3)$ lattice results away from the chiral limit.
Your results might enable phenomenologists to sharpen their predictions (and improve them).
To enable this, make sure your results are packaged in a broadly useful way: include results for
physical quantities even away from the physical point of QCD, and include dimensionless ratios
that can be used to set the scale in a variety of ways.

Second, phenomenology makes heavy use of models.
``Model'' is a heretical word to the lattice community, but that is not so, outside it.
 It can be useful to present lattice results
in a way which allows easy comparison with models, rather than as stand - alone results.  The
KSRF relations discussed above, which can be obtained from models of vector meson dominance, are a good example.


\begin{acknowledgments}
We would like to thank
John Bulava,
Jim Halverson,
and
Yuhsin Tsai
 for conversations and correspondence.
This work was supported in part by the U.S. Department of Energy under grant DE-SC0010005.
 Brookhaven National Laboratory is supported by the U. S. Department of Energy under contract DE-SC0012704.
\end{acknowledgments}



\end{document}